# Positionally precise functionalization of shallow luminescent centers through Forster Resonant Energy Transfer (FRET) driven surface photochemistry.

Alexei Goun, Herschel Rabitz

agoun@princeton.edu

Department of Chemistry, Princeton University, Princeton, NJ 08544

Abstract

Paramagnetic luminescent impurities in solids such as Nitrogen-Vacancy (NV) centers in diamond represent a promising and versatile platform for the development of a wide range of chemical and biological sensors. This goal can be accomplished by the placement of chemically active, spin (electron or nuclear) labeled moiety on the surface in the close vicinity of a shallow single paramagnetic center. In this paper, we demonstrate that the Forster Resonant Excitation Transfer driven process where luminescent center plays a role of excitation donor can accomplish such a goal with high chemical efficiency, positionally precise and a scalable manner. We obtain the probability distribution function of the sensing group position relative to the luminescent center (NV center) and demonstrate that the functionalization position uncertainty is equal to the luminescent center deposition depth. The efficiency of the FRET process is analyzed as a function of the luminescent center deposition depth and the density of the excitation acceptor sites. Employing the geometric information of the surface-functionalized sensing groups, we obtain the probability distribution function of the energy gap in the spectrum of a spin-based detection system. This information allows us to estimate the single-molecule detection capabilities of the proposed system.

**Introduction.**

The recent rapid development of technology around single luminescent defects in solids opens a wide range of new capabilities in quantum information science, single-molecule structure, and dynamics analysis as well as allows the creation of a whole new range of biochemical detection platforms [1]. The key to these capabilities is the long spin coherence time of luminescent paramagnetic centers such as Nitrogen-Vacancy (NV) sites in isotopically pure hosts (diamond). There is a wide variety of luminescent paramagnetic impurities NV centers in diamond, Magnesium in GaN [2], Phosphorous in Silicon, and other systems. The discussion below can be applied to any of these systems, with adjustment in the surface chemistry. We would discuss our approach within the context of NV centers in diamond.

In the detection assays, the analyte molecule finds its selective binding partner, and the docking event is reported to the observer. This final reporting is typically done through optical emission detection, for example, the analyte molecule disturbing the excitation transfer process between donor and acceptor moieties and increasing the emission of the donor. Despite significant progress, optical emission-based detection essays face significant challenges as every new variant typically necessitates a novel molecular design that is expensive and time-consuming. Chromophores also have a strong tendency for photodegradation and many assay configurations are not scalable to simultaneous detection of a large number of biological and chemical targets. Thus, an essay that is label-free, operating on a broad range of chemical and biological targets (analytes) is highly desirable in heal care, biomedical research, and defense applications.

Spin-based biological and chemical detection systems offer a novel, promising approach allowing highly integrated assays. The technique of spin labeling of proteins is well developed and allowed a broad range of measurements of proteins dynamics and structure. Spin-labeled antibodies are also regularly utilized [3] in structural and dynamic measurements. The spin-label can be either covalently attached to the protein or form a thermodynamically stable complex [4]. Detection of the single spin-label, attached to the surface of the protein with the help of the NV center was recently demonstrated [5]. We suggest the chemical and biological detection approach through the use of spin-labeled molecular sensors. The proposed operation of the spin-based molecular sensor is shown in Figure 1. The detector is represented by the isotopically pure CVD-grown diamond chip, with a single layer of NV- centers deposited in the vicinity of the surface using the delta-doping approach. The density of NV- centers in the layer is low enough to perming individual optical access and detection to individual centers. For the emission wavelength of 675nm, the Abbe resolution limit $\lambda/2NA$ implies the possibility to integrate up to ~$10^8$ detection centers per cm$^2$ of the chip. Large separation between NV centers within the monolayer eliminates the broadening of spin resonance transition due to magnetic interactions among centers. Spin-based optical sensors are added to the chip through the route described in the next section. The liquid, containing the mixture of analytes is deposited on the surface of the chip with the help of microfluidic components. Analytes in the mixture interact with a surface-bound array of sensors and find the complementary targets through the diffusion-limited reaction.

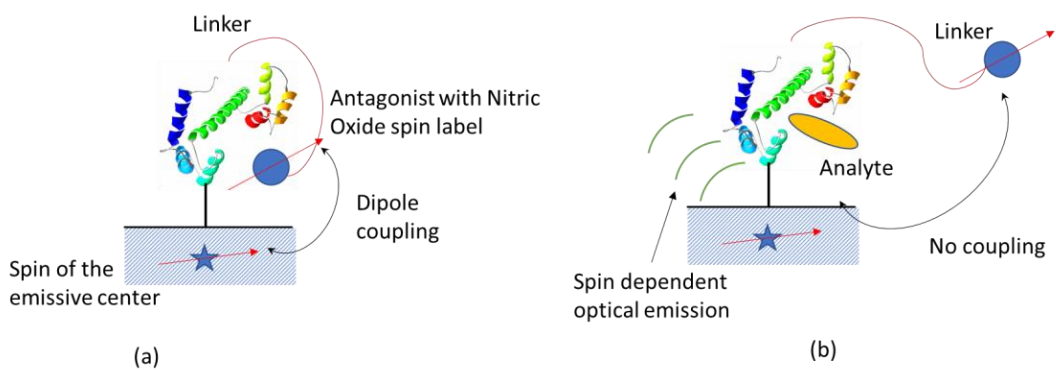

Figure 1. Optically Detected-Electron Paramagnetic Resonance (OD-EPR) spin coupling-based ligand binding sensor. (a) The spin-labeled antagonist molecule is attached to the binding domain (antibody), no Analyte molecule is present. Due to the proximity of the spin-label to the NV center, there is a strong magnetic interaction, resulting in a large splitting in the EPR spectrum (b) The detection system after the attachment of the analyte molecule. The Analyte molecule

displaces the spin-labeled Antagonist, significantly reducing the spin-spin magnetic interaction, and causing the reduction of the splitting in the EPR spectrum.

The core detection technique is the Optically Detected-Electron Paramagnetic Resonance (OD-EPR) [6]. The emission of the NV center in diamond is dependent on its spin state, which allows measurement of the EPR spectrum of its environment through optical detection. Our proposed sensing system is composed of two components, the surface chemically/biologically sensing domain with spin-labeled antagonist molecule and the luminescent NV center that serves as a reporter for the spin state of the system. In this case, the sensing and optical reporting are separated, the sensing domain is not subject to photodegradation, and a broad range of sensing domains can be easily constructed. The attachment of the analyte molecule will cause a geometric change in the binding domain, affecting the coupling between the spin-labeled antagonist and the NV center.

The key to detection and in general quantum information processing capabilities is the long spin coherence time of NV centers. To ensure such a long time, the NV center must be placed sufficiently deep within the diamond, at the same time the center must be close enough to the surface to allow the interaction with external electron and nuclear spins. The ability to satisfy these conditions, together with a high degree of photostability gives NV centers their unique capability. It is shown that the useful deposition depth of NV is 3-20nm. By utilizing a delta-doping scheme this depth can be tightly controlled [7]. Spins at smaller deposition depths suffer from an increased rate of dephasing. Spins at large deposition depth do not couple efficiently to spins on the surface.

To evaluate the detection capability of the proposed chemical and biological sensor, one needs to understand the geometric distribution of the sensor position in its two states (analyte present/analyte absent). We obtain such a geometry distribution through the analysis of the sensing domains attachment process that is driven by the combination of the photochemical reaction and FRET excitation transfer process.

**Positionally Accurate Functionalization through Forster Excitation Transfer Driven Surface Photochemistry.**

The critical capability that is still missing is the positionally precise functionalization of the surface to bring the spin-based sensor into the vicinity of the buried NV center. There is no chemical information on the surface of the diamond to indicate the presence of the NV center. In sensing applications, one needs to place the analyte molecule within ~ 5nm of the NV center to enable the detection. In single-molecule NMR experiments, the positioning requirements are even higher due to the weakness of nuclear-electron spin coupling.

At present there are two methods of surface positioning, both are not satisfactory. In the first (stochastic), the sample is spread uniformly over the surface and if the density is high enough, then the molecule of interest will be near the NV center. This approach suffers from low sensitivity as most analyte molecules are bound to sensing domains that are too far away from NV centers. Another approach is to use scanning microscopy to locate the NV center and then attempt to place a molecule of interest in its vicinity. This approach is not scalable and suffers from poor positioning accuracy.

We offer an alternative approach that is based on the combination of photochemical uncaging of surface-bound protective groups and Forster Resonant Energy Transfer (FRET) from the NV center to these groups. We determine the efficiency of the excitation transfer as a function of the NV center deposition depth, the surface density of the excitation accepting site, and the excitation transfer distance. We demonstrate that the proposed technique is efficient and spatially precise. The FRET-driven process should be contrasted with the emission – absorption process with a former is a coherent coupling through the electromagnetic field. The emission reabsorption process is limited in its accuracy to the wavelength of NV emission and is inefficient due to the low optical density of the absorbing media. FRET drive process is free of these shortcomings due to the much stronger dipole-dipole coupling with a donor-acceptor coupling efficiency approaching 100% given a short distance between them. The suggested functionalization processes can be accomplished through several routes such as FRET-driven deprotection of surface-bound photolabile groups (*Figure 2*) and a FRET-driven excitation of photoinitiator in a photoresist (*Figure 3*). As an additional route and FRET-driven photoredox catalyzed surface chemistry can be utilized [8], but the diffusion transport of the redox reaction might limit the spatial resolution of the functionalization.

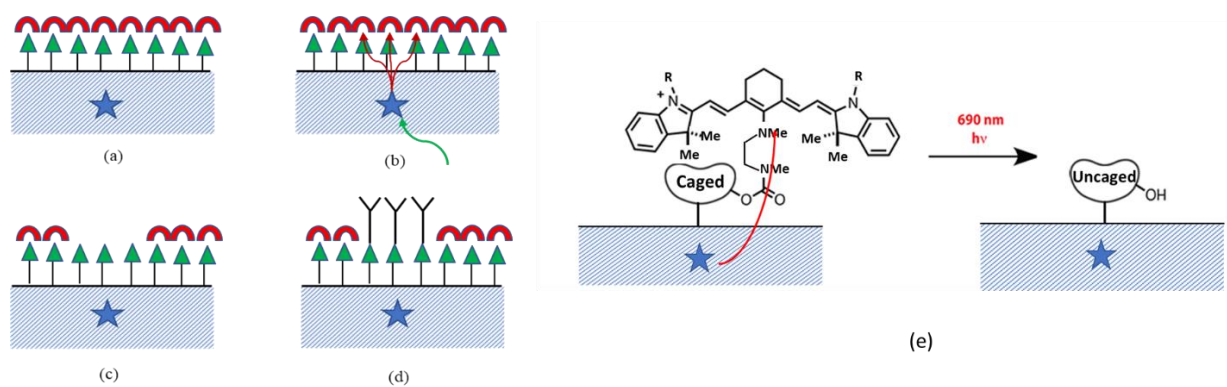

Figure 2. FRET-driven surface functionalization through excitation transfer to surface-bound photolabile groups. a) Coating of diamond with photo-uncaging protective groups b) Optical Excitation of NV Center and FRET transfer of the electronic excitation to surface-bound groups c) Uncaging of the surface protective functional groups d) Attachment of antibody-based sensing domains to the exposed functional groups e) Photoprotective groups based on the dissociating near-IR cyanine dye.

The first suggested approach (Figure 2) contains four steps. In the first step of the deposition process, shown in Figure 2a, the surface of the diamond sample is covered with a monolayer, containing the photo-uncaging protective group (i.e. the functional group that dissociates under the photoexcitation). Such a process can be accomplished utilizing the UV grafting of CVD grown isotopically pure diamond [9, 10], where the attachment of the amine functional group was demonstrated. Another route to couple the functional groups to the diamond surface is through the use of carboxylic groups on the diamond surface. To activate the surface carboxylic groups a water-soluble mixture of 1-ethyl-3-(3-dimethylamino)propyl)carbodiimide (EDC) and N-hydroxy-sulfosuccinimide (NHS) [11]. Non-covalent, physisorbed coatings are also possible, although they do not provide the resilience, expected from detection systems. The photoprotected caged group can be attached to the surface first, and the protective photosensitive group can be attached later. The absorption spectrum of such a photoprotective group lays in the red to the near-IR portion of the spectrum and strongly overlaps with the emission spectrum of NV-center. An example of such a photo-uncaging group can be found in [12] and it is based on near-IR cyanine dye. At the second step shown in *Figure 2*b, the sample is illuminated with optical radiation, within the region of NV center absorption (from 450nm to 600nm). Once photoexcited, the NV center undergoes Forster Resonant Excitation Transfer (FRET) to the optical chromophore of the photoprotective group. FRET transfer is a very short range (4nm, typically) and is highly efficient within that range. At the third step, shown in *Figure 2*c the excited protective groups dissociate, leaving the functional group sterically available for the attachment process. In the fourth step, shown in *Figure 2*d the available functional groups attach the molecule of interest such as a spin-labeled antibody.

The second suggested route is illustrated in *Figure 3*. This route provides a colocalization of the luminescent center with surface-bound functional groups through the FRET process to the red light-sensitive photoresist [13]. The photoresist is the polymer system that can provide exceeding thin coatings down to few nanometers. The photoresist contains a light-absorbing photoinitiator molecule that controls the initiation of the polymerization. The photoinitiator serves as an acceptor of the FRET excitation transfer process with a donor being the luminescent center.

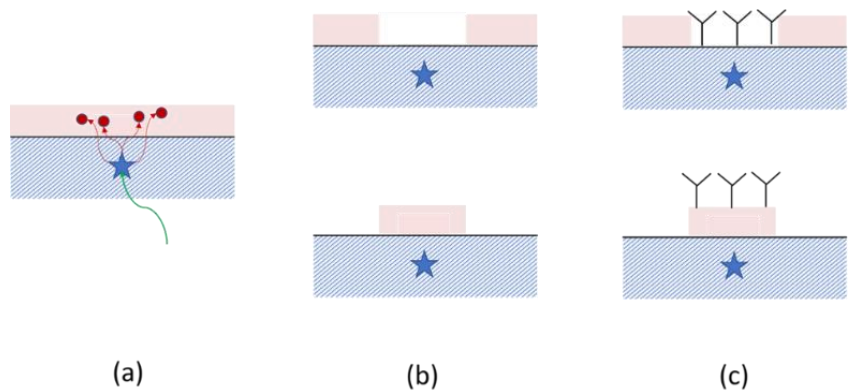

(a)      (b)      (c)

Figure 3. Positionally accurate functionalization through excitation transfer to the red light-sensitive photoinitiator of the photoresist. a) Optical excitation of the luminescent center, and excitation transfer to the photoinitiator of photoresist b) Removal of the exposed positive photoresist (top) and unexposed negative photoresist (bottom) c) Functionalization of the exposed substrate surface (top), functionalization of the remaining photoresist (bottom).

In this section the efficiency of the attachment process and the spatial distribution of attachment sites will be analyzed, corresponding to the process illustrated in *Figure 2*. The spectral properties of near-IR photo uncaging groups based on cyanine dyes will be utilized in the analysis [12].

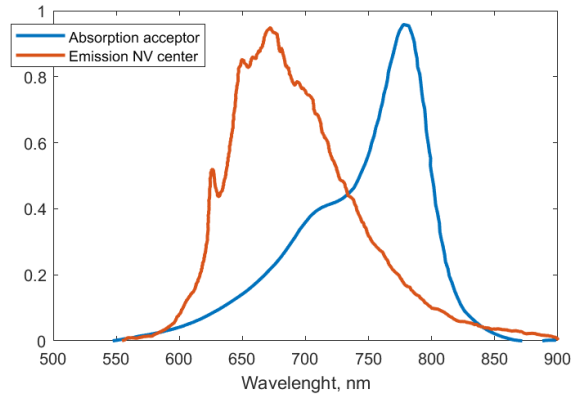

Figure 4. The emission spectrum of the NV center (red), the absorption spectrum of excitation acceptor (IR-780). IR-780 is the light-sensitive component of photo-uncaging, a protective group from [12]

Utilizing the emission spectrum of the NV center [14], the absorption spectrum of IR-780 based functional group [15], the excitation transfer distance is given by [16]

$$R_0^6 = \frac{9000(ln10)k^2 Q_D}{N_A 128\pi^5 n^4} \cdot \frac{\int_0^\infty \varepsilon(\nu) F_D(\nu) \nu^4 d\nu}{\int_0^\infty F_D(\nu)\, d\nu} \quad (1)$$

Given the refractive index of diamond of 2.4, the FRET distance $R_0$ is 4.3nm.

The relative geometry of the NV center with acceptor sites is illustrated in *Figure 5*.

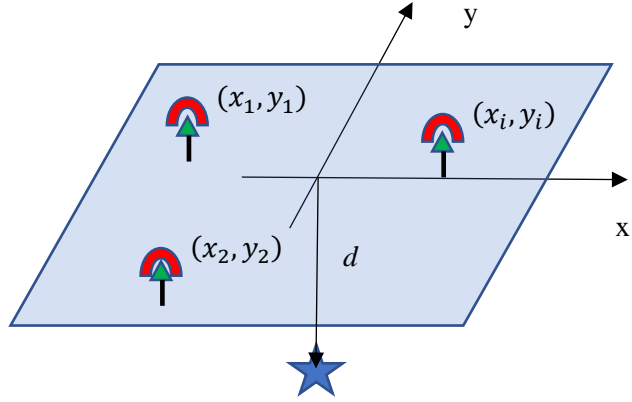

Figure 5. The relative positions of the NV center and surface photo-uncaging groups, $(x_i, y_i)$ is the position of the $i^{th}$ surface excitation accepting group, $d$ is the deposition depth of the NV center.

The optical excitation is transferred from a single luminescent center to the distribution of acceptors on the surface. The excitation of the donor is transmitted to a collection of acceptors with random distributions of positions. The evolution donor excited state probability for a given distribution of acceptors $P_{ex}$ is given by.

$$\frac{dP_{ex}}{dt} = -\left(\frac{1}{\tau_{DL}} + \sum_{i=1}^{N} k_f(x_i, y_i)\right) P_{ex} \tag{2}$$

Where $\tau_{DL}$ is the excited state lifetime of the donor molecule and $k_f(x_i, y_i)$ is the excitation transfer rate from donor to acceptor at $(x_i, y_i)$.

$$k_f(x_i, y_i) = \frac{1}{\tau_{DL}} \frac{R_0^6}{(d^2 + x_i^2 + y_i^2)^3} \tag{3}$$

Once the excitation is transmitted to the acceptor molecules it can either dissipate or cause the dissociation of the protective group. The evolution of the acceptor molecule $i$ at the position $(x_i, y_i)$ is described by the following equation.

$$\frac{dP_{ET}^i}{dt} = -\left(\frac{1}{\tau_{AL}} + k_D\right) P_{ET}^i + k_f(x_i, y_i) P_{ex} \tag{4}$$

Where $\tau_{EA}$ is the excited state lifetime of the acceptor molecule, $k_D$ is the rate of dissociation (uncaging). The evolution of Eq (2) for $P_{ex}$ depends on the random spatial distribution of acceptors. The averaged thermodynamic limit of excited-state evolution is given by [17, 18]

$$P_{ex}(t) = exp\left(-\frac{t}{\tau_{DL}}\right) exp\left(-2\pi C \int_0^\infty [1 - S(t,r)] r dr\right), \tag{5}$$

where $S(t, r)$ is the survival probability of a single donor in the presence of a single acceptor pair, $C$ is the acceptor surface concentration.

$$S(t,r) = exp\left(-\frac{t}{\tau_{DL}}\frac{R_0^6}{(d^2+r^2)^3}\right) \qquad (6)$$

The survival probability of the donor excited state after some simple rearrangements:

$$P_{ex}(t) = exp\left(-\frac{t}{\tau}\right)exp\left(-\pi C d^2 \int_0^\infty \left[1 - exp\left(-\frac{t}{\tau_L}\left(\frac{R_0}{d}\right)^6 \frac{1}{(1+y)^3}\right)\right]dy\right) \qquad (7)$$

Let us consider the behavior of the "form factor" integral

$$f(\alpha) = \int_0^\infty \left[1 - exp\left(-\alpha \frac{1}{(1+y)^3}\right)\right]dy, \qquad \alpha = \frac{t}{\tau_L}\left(\frac{R_0}{d}\right)^6 \qquad (8)$$

The plot of the function $f(\alpha)$ is shown in *Figure 6*a below. This function allows us to compute the decay of the donor excited state population for the entire range of the concentrations. The decay of the excited state population of the NV center in the presence of different concentrations of surface acceptors is shown in *Figure 6*b. Nitrogen vacancy excited state lifetime $\tau_L = 23ns$, [19] the deposition depth is 3nm. It is easy to see that the presence of the surface acceptor groups rapidly reduces the excited state lifetime and the quantum yield of NV center luminescence. The reduction in the emission is due to the appearance of the excitation transfer channel, and it can be used to quantify its efficiency. From the numerical analysis, it follows that for the photo-uncaging group, described in [12] and NV center deposited at 3nm depth, 50% chance of excitation transfer is at the surface acceptor density 0.02 groups/nm² for deposition depth of 5nm, the surface density is 0.07 groups/nm²; for 10nm is at 1 acceptor/nm². *Figure 6*c shows the quantum yield of the excitation transfer to surface uncaging groups as a function of acceptor concentrations for different NV center deposition depths. It is easy to see that even for the largest deposition depths, there is an appreciable probability of excitation transfer to surface groups.

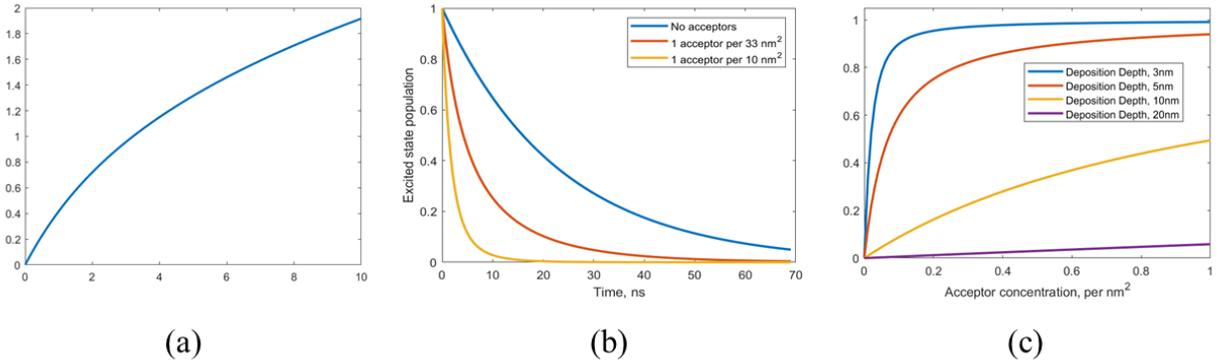

(a)         (b)         (c)

Figure 6. a. The "form factor" of the excitation decay. b. Decay of the excited state of NV center. c. Probability of transfer to the surface group as a function of surface acceptor group concentration.

Let us evaluate the statistical characteristics of the surface distribution of available binding sites once the photouncaging group dissociated following the optical excitation. Using Eq. (4) for given acceptor site $i$ we obtain the following probability of excitation

$$P_{ET}^i = k_f(x_i, y_i) \int_0^t P_{ex}(t_1) exp\left(-\left(\frac{1}{\tau_{AL}} + k_D\right)(t - t_1)\right) dt_1 \quad (9)$$

The important property of Eq. (9) is that the spatial and temporal dependences of the probability factorize, and all the acceptor sites have identical time evolution of un-caging probability. Thus, the spatial distribution of the attachment probability is proportional to $\sim k_f(x, y)$. The properly normalized probability distribution of the attachment sites is then given by Eq (10). The probability distribution is illustrated in Figure 7.

$$\rho(r) = \frac{2d^4}{\pi} \frac{1}{(d^2 + r^2)^3} \quad (10)$$

The dispersion of distances of the accepting cites from the origin is given by

$$\langle r^2 \rangle = \int_0^\infty \rho(r) r^2 \pi r dr = \int_0^\infty \frac{4d^4}{\pi} \frac{1}{(d^2 + r^2)^3} r^2 \pi r dr = d^2 \quad (11)$$

The shallower the NV center deposition, the tighter the lateral coupling becomes with a spin sensor.

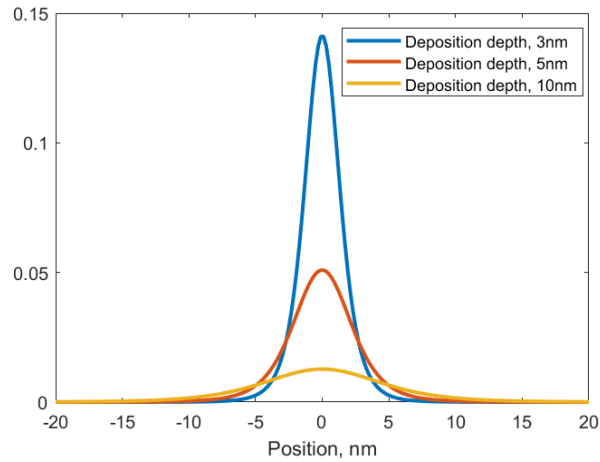

Figure 7. The probability distribution of attachment sites for different depths of deposition of NV centers.

The distribution of distances, given by Eq. (10) provides a statistical description of the geometric ensemble of the sensing region attachment in the vicinity of the reporting NV-center and allows us to assess sensor performance, its Receiver Operating Curve (ROC curve).

**Chemical and Biological Sensor Based on Optically Detected Electron Spin Resonance.**

Now, as we have obtained the statistical description of the ensemble of possible attachment geometries, we can analyze the statistical properties of spin-based detectors, illustrated in Figure 1. The distribution of the sensing site positions around the NV center creates a distribution of spin-spin couplings and consequently the distribution of possible EPR spectra.

The Hamiltonian of the NV center is given by

$$\widehat{H}_{NV} = \hbar D \left[\hat{S}_z^2 - \frac{2}{3}\right] + \hbar \gamma \vec{B} \cdot \hat{S} \tag{12}$$

Where $\hat{S}$ are the spin operators, $D$ is the "zero-field split" parameter, equal to 2.87 GHz, $\vec{B}$ is the external magnetic field, and $\gamma$ is the electron's gyromagnetic ratio. Once coupled to the external spin 1/2 label based sensor, the overall spin Hamiltonian becomes.

$$\widehat{H}_{NV-s} = \left(\hbar D \left[\hat{S}_z^2 - \frac{2}{3}\right] + \hbar \gamma \vec{B} \cdot \hat{S}\right)_{NV} \otimes 1_s + 1_{NV} \otimes \hbar \gamma \vec{B} \cdot \hat{\sigma} \tag{13}$$
$$+ g^2 \beta^2 \left[\frac{\hat{S} \cdot \hat{\sigma}}{\bar{r}^3} - \frac{3(\hat{S} \cdot \vec{r})(\hat{\sigma} \cdot \vec{r})}{\bar{r}^5}\right]$$

Where $\hat{\sigma}$ is the spin operator of the label, $\bar{r}$ is the radius vector from the NV center to the spin-label. $g^2\beta^2$ determines the scale of the magnetic coupling between the NV center and the spin-label, its numerical value is $\hbar \cdot 0.33$GHz. The attachment of the analyte molecule to the sensor will displace the spin-labeled antagonist molecule. The initial spin-labeled attached configuration and the final spin-labeled displaced by the analyte are distinct random spatial distribution. In both configurations the spin-label occupies a certain volume of the configuration space (position and orientation) and undergoes a diffusion which can result in complex modification of line shapes such as motional narrowing, potentially significantly improving the resolution of the ESR measurement. To incorporate the effect of diffusion and steric limitations on all possible configurations, a complex molecular dynamic simulation coupled to the spin dynamics is required. We would estimate the performance of the sensor by neglecting the effect of diffusional motion at each step of the ESR measurement and assume that positions of the spin-label are fixed within a sphere of uncertainty in both sensor inactive (analyte unattached, item 3 of *Figure 8*) and sensor active (analyte attached, item 4 of *Figure 8*) configurations. Additional detection uncertainty is introduced by the precise position of the attachment of the sensor to the diamond surface (item 2 of *Figure 8*) with a distribution of distances given by Equation (10).

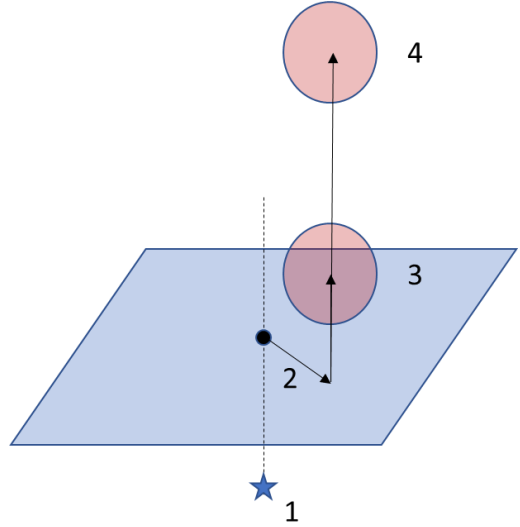

Figure 8. The mutual spatial configuration of the NV center (1) and the spin-label in the sensor analyte not bound (OFF) state (3) and the analyte bound (ON) state.

Every random spatial configuration of the NV-center and the spin-label is generated by the random attachment point with probability density given by Equation (10) and the uniform random distribution of the spin-label within the sphere of uncertainty. We would assume that the detection parameter that separates the sensor ON and OFF states are the heights of the spin-label above the diamond surface. For every configuration from the random set of possible configurations, the energy splitting $\Delta E$ between transitions $m_{S=0} \to m_{S=+1}$ and $m_{S=0} \to m_{S=-1}$ is calculated. Such a parameter serves as a sensitive measure for NV-center – spin-label coupling and will indicate the geometry and consequently the state of the sensor. The ensemble of geometries will then induce a probability distribution function of $\Delta E$. The variation of the probability distribution of $\Delta E$ with the heigh of spin label above the diamond surface (or with other variation of the geometry and dynamics) forms the basis of the detection and allows us to compute the statistical characteristics of the detection system (receiver operating characteristic, ROC curve). The change in the probability density function with the sensor-surface separation is shown in *Figure 9*. The further the separation of the spin-label from the surface, the smaller is the energy separation. One can notice the spiking of the probability density function around zero energy, corresponding to configurations with the largest separation from NV center to spin-label, occupying a large portion of the state space. There is also a sharp cut-off in the region of high interaction, corresponding to the configuration of the closest approach of the NV center and sensing domain.

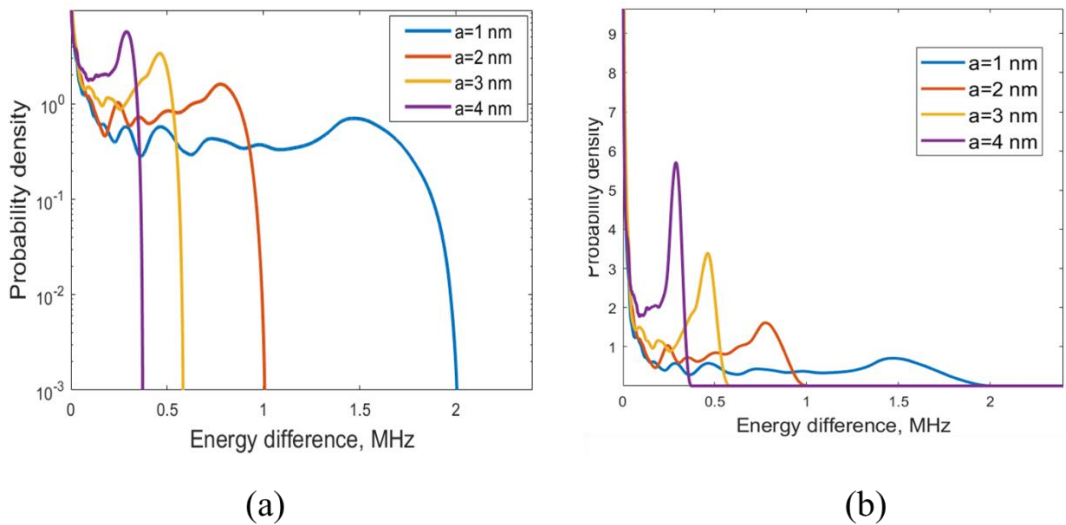

(a)                                                       (b)

Figure 9. The probability distribution of the frequency separation in the optically detected ESR spectrum of the NV center coupled to the spin-label for different distances of spin-label above the diamond surface. The NV center deposition depth is assumed to be equal to 3nm and the sphere of uncertainty is assumed to have a diameter of 0.5 nm.

Once the statistical description of the measurement is obtained, the ROC curve can be obtained. We would assume that thetate sensor OFF corresponds to the position of 1nm above the surface. The ROC curves for the several ON positions of the sensor are shown in Figure 10 and Figure 11. As expected, the performance of the sensor is improved with higher geometric contrast of ON and OFF states. It should be noted that the sensor exhibits a very low False Alarm Rate and an appreciable probability of detection.

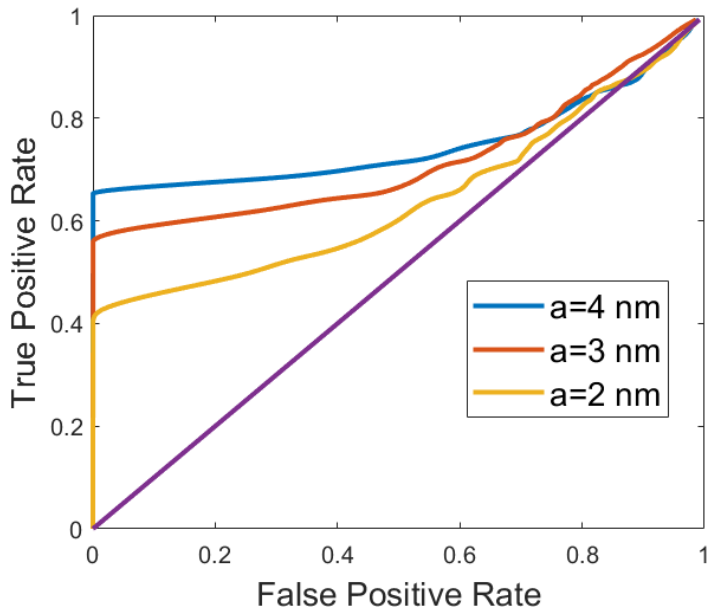

Figure 10. The ROC curve of the spin sensor-NV center for various positions (displacement of spin-label above the diamond surface) of the ON state. The position of the OFF state is 1nm above the surface.

The performance of the sensor is significantly improved if the deposition depth is shallow as it offers stronger coupling to the spin-label and higher contrast in the energy difference variation. The figure below illustrates the ROC curves for three different deposition depths.

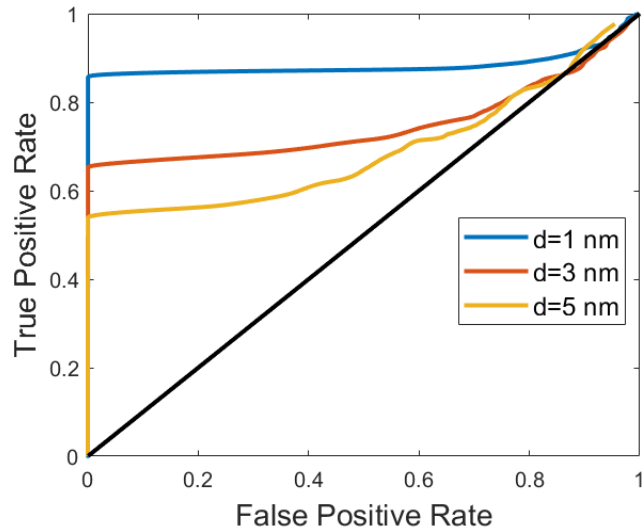

Figure 11. Receiver operating characteristic of the spin sensor – NV center for different deposition depths. The position of the OFF state is 1nm above the surface, the position of the ON state is 1nm above the surface, the diameter of uncertainty is 0.5nm.

**Conclusion.**

In this paper, we proposed the route for positionally precise functionalization of shallow luminescent defects in solids. The expected performance of such a procedure was analyzed using parameters of NV centers in diamond. Our analysis indicates high positional accuracy of such a functionalization, with a standard deviation of the functional group position, is equal to the deposition depth of the NV center. The scheme and analysis proposed in the paper can be extended to a broad range of luminescent centers in solids. The FRET-driven route for surface functionalization can be extended to other optoelectronic components such as single quantum dot emitters [20].

We have also analyzed the coupling of the NV center to the external spin-labeled sensing group and demonstrated the potential low false alarm rate of such a system. A large number of single luminescent sensors can be placed on the chip, potentially allowing a very versatile, cost-effective chemical and biological sensing system.

The presented analysis did not take into account the influence of the hyperfine splitting due to electron-nuclear spin interaction. These interactions would make the analysis of sensor operation more complex and are dependent on the hyperfine structure of molecules participating in the sensing domain. Adaptive algorithms such as Optimal Dynamic Discrimination [21] can further improve the fidelity of sensing operations. These broadening mechanisms can also be significantly reduced or eliminated by utilizing nuclear spin-free isotopes for chemical and biological sensing domains, as well as the deuterated solvent.

Besides the detection applications, the positionally accurate functionalization of the NV centers will significantly extend the capability for sing molecule NMR and EPR measurements thus providing structural and functional insights into the operation of biomolecules. The integration of the positionally accurate functionalization with atomic force microscopy will allow the real-time observation of the restructuring of complex spin states during the operation of complex proteins.

**References.**